\documentclass{llncs}

\usepackage{amsbsy}
\usepackage{amssymb}
\usepackage{mathrsfs}
\usepackage{latexsym}
\usepackage{enumerate}
\usepackage{amsmath}
\usepackage{amsfonts}
\usepackage{float}
\usepackage{wrapfig}
\usepackage{lipsum}
\usepackage{caption}

\newtheorem{lemma1}{Lemma}[section]
\newtheorem{theorem1}{Theorem}[section]
\newcommand{\qedsymb}{\hfill{\rule{2mm}{2mm}}}
\newenvironment{proof1}{\begin{trivlist}
\item[\hspace{\labelsep}{\bf\noindent Proof: }]
}{\qedsymb\end{trivlist}}

\floatname{algorithm}{Protocol}

\floatstyle{plain}
\newfloat{myalgo}{tbhp}{mya}

{\begin{myalgo}[#1]
\centering
\begin{minipage}{6.0in}
\begin{algorithm}[H]}%
{\end{algorithm}
\end{minipage}
\end{myalgo}}

\begin{document}

\title{\LARGE\bf The complexity of resolving conflicts on MAC}

\author{Shailesh Vaya}
\institute{Xerox Research Centre India
\newline  Bangalore, India - 560048
\newline Email: shailesh.vaya@gmail.com}

\maketitle
\abstract{
\noindent
  We consider the fundamental problem of multiple stations competing to transmit on a multiple access channel (MAC). We are given $n$ stations out of which at most $d$ are active and intend to transmit a message to other stations using MAC. All stations are assumed to be synchronized according to a time clock. If $l$ stations node transmit in the same round, then the MAC provides the feedback whether $l=0$, $l=2$ (collision occurred) or $l=1$. When $l=1$, then a single station is indeed able to successfully transmit a message, which is received by all other nodes. For the above problem the active stations have to schedule their transmissions so that they can singly, transmit their messages on MAC, based only on the feedback received from the MAC in previous round.

	For the above problem it was shown in [Greenberg, Winograd, {\em A Lower bound on the Time Needed in the Worst Case to Resolve Conflicts Deterministically in Multiple Access Channels}, Journal of ACM 1985] that every deterministic adaptive algorithm should take $\Omega(d (\lg n)/(\lg d))$ rounds in the worst case. The fastest known deterministic adaptive algorithm requires $O(d \lg n)$ rounds. The gap between the upper and lower bound is $O(\lg d)$ round. It is substantial for most values of $d$: When $d = $ constant and $d \in O(n^{\epsilon})$ (for any constant $\epsilon \leq 1$, the lower bound is respectively $O(\lg n)$ and $O(n)$, which is trivial in both cases. Nevertheless, the above lower bound is interesting indeed when $d \in$ poly($\lg n$). In this work, we present a novel counting argument to prove a tight lower bound of $\Omega(d \lg n)$ rounds for all deterministic, adaptive algorithms, closing this long standing open question.
}


\setcounter{page}{1}
\section{Introduction}
\label{sec:adaptive}
  In telecommunications and computer networks, a multiple access channel (MAC) allows several terminals connected to the same multi-point transmission medium to transmit over it and thus share its capacity. There are many examples of such networks like wireless networks, bus networks, ring networks, hub networks and half-duplex point-to-point links \cite{cj1}, \cite{cj2}, \cite{G}, \cite{GL}, \cite{M}, \cite{TM}. The following model of communication on this channel is commonly adapted for theoretical studies: There is an ensemble of $n$ communication stations with labels $\{1, 2, 3, \dots \}$. $d$ of these stations are live and intend to transmit a message to the other $n-1$ stations. All stations are synchronized with respect to a centralized clock and one or more stations can attempt to transmit a message in any round. Following are the possible consequences when $l$ stations attempt to transmit a message on MAC in a given round
\begin{enumerate}
\item{} When $l = 0$ no packets are transmitted.
\item{} When $l = 1$ a single station is able to successfully transmit its message to every other station (along with its label).
\item{} When $l \geq 2$ transmissions interfere with one another so that none of the packets are successfully transmitted; More over collision is detected by the stations.
\end{enumerate}

  An algorithm for the above problem determines how to schedule the retransmissions of the stations on the MAC so that each of the $d$ stations can successfully transmit on the channel once. All stations can determine whether to transmit or not based in a given round, depending on what they hear on the MAC in the previous round. These type of algorithms are called adaptive. On the other hand in Non-adaptive algorithms the transmission schedule of the stations that determines whether a station is to transmit in a given round or not, is pre-determined and may only depend on the value of $n, d$ and the label of the station itself. In this work, we study adaptive, deterministic algorithms for conflict resolution.

  Komlos and Greenberg present a non-adaptive protocol that takes $O(d^2 \lg n)$ rounds for the conflict resolution problem in \cite{KG}. Capetanakis \cite{cj1} \cite{cj2}, Hayes \cite{H}, Tsybakov and Mikhailov \cite{TM} independently found an adaptive deterministic \textit{tree algorithm} to resolve conflicts, which runs in $\Theta(d + d \lg (n/d))$ rounds in the worst case. Greenberg and Winograd \cite{GW} present a very interesting information theoretic lower bound of $\Omega(d (\lg n)/(\lg d))$ rounds for every adaptive, deterministic algorithm for the problem. The proof uses an adversary to act as a foil against the algorithm, finding a conflict that is particularly hard to resolve.

  On the other hand, a probabilistic version of the tree algorithm resolves conflict without error in expected time approximately $2.885 d$ no matter what the identities of the $d$ conflicting stations are \cite{M}, \cite{TM}, which can be further improved to $2.14 d + O(\lg d)$. Willard \cite{W} presented tight lower bound on the expected running time of any probabilistic algorithm for obtaining a single successful transmission, after a conflict involving $2$ or more stations.

\subsection{Recent developments}
\label{relatedworks}
  Besides communication networks, the elegant Query game type problem described above, appears in other areas of Computer Science like Combinatorial group testing \cite{DH}, single hop radio networks (called the selection problem) \cite{K}, Straggler identification \cite{G} etc.. The information theoretic lower bound presented in \cite{GW} also appears in \cite{DH} [Du, Hwang, Chapter 5, Combinatorial Group Testing and its applications (Discrete Applied Mathematics - vol 3), page 94 - 96].

	Mutiple access channels have been extensively studied in the literature - \cite{K}, \cite{CGKP}, \cite{E}, \cite{CR}, \cite{CKR}, \cite{CKL}, \cite{ACR}, \cite{BCHKL},\cite{MPR}, \cite{MPS}, \cite{HLPW}, \cite{GH}, \cite{AMM},\cite{AM} - all discuss one or more variants of the conflict resolution problem in multiple access channels.

\section{Preliminaries}
\label{preliminaries}
  We introduce a few terms and notations to formalize the problem of conflict resolution in multiple access channels. We also state relevant assumptions about the set up.

\begin{definition}
\label{qgame}
  A \textbf{Q-game} is the following conflict resolution problem for multiple access channels. There is a set of $N$ stations numbered $1, 2, \dots, n$ which can transmit data packets on a multi-access channel at steps numbered $1, 2, 3,\dots$. The goal is to get all stations belonging to subset \textbf{S}, $|S| = $\textbf{d} transmit their messages to the remaining $n - 1$ stations, given the following characteristic of the transmission process on the multiple access channel (MAC). If in a particular round $l$ stations transmit, then the result of that depends on $l$:	
\begin{enumerate}
\item{} If $l = 0$, no packets are transmitted.
\item{} If $l = 1$, a station is able to successfully transmit its packets to every station in $N$.
\item{} If $l \geq 2$, then collision occurs, so that none of the packets are transmitted successfully. Stations have a collision detection mechanism by which they are able to infer whether $2$ or more stations transmitted in a given round.
\end{enumerate}
\end{definition}

  All stations are assumed to be synchronized with respect to a centralized clock. All stations certainly receive the feedback whether $0, 1$ or $2+$ stations transmitted in the previous step. In fact, the transmitting stations can attach their labels along with their messages. When a single station is able to successfully transmit by itself, the other stations learn its label besides its message. Thus, there are $d + 2$ possibile outcomes of transmissions in any given round.

  The setting of the above problem is very similar to the following classical problem in combinatorial group testing \cite{DH}. We are given a set $N$ of $n$ objects, $N = \{x_1, x_2, \dots, x_n\}$, out of which $d$ are defective. The task is to find the $d$ defective objects using as few tests as possible given certain characteristic of the querying process. In any test with querying set $T_s$: (a) If there are $0$ defective items in $T_s$, then this fact is learnt (b) If there are $|T_s| \geq 2$ defective items, then it is learnt that $\geq 2$ defective objects exist in $T_s$ (c) If however $T_s$ has exactly one defective item, then its identity is learnt.

\section{Lower bound of $\Omega(d \lg_2 n)$ transmission rounds for Q-games}
\label{sec:lower}
  In \cite{GW}, the authors give an information theoretic lower bound of $\Omega(d \lg_d n)$ rounds for $d$ live stations to make  transmissions singly when there are $n$ total channels. To give a flavor of the new lower bound, we sketch an alternate proof of the older result.

\subsection{An alternate sketch for lower bound of $\Omega(d \lg n/\lg d)$ rounds for Q-games}
\label{alternate}
  We provide a combinatorial argument for the older lower bound.
  Fix any scheme $T$ for the \textbf{Q}-game that makes $t$ queries in the worst case. Consider two distinct subsets of live channels $X_1$ and $X_2$ for which $X_1 \neq X_2$. Then, it must be the case that the output of some query from the sequence of at most $t$ queries should be different. Or else the sequence of queries would not result in making distinguishing transmissions corresponding to subset of live channels $X_1$ and $X_2$. Thus, any querying scheme $T$ should result in at least $\binom{n}{d}$ distinct outcomes to the sequence of queries. Now note that the maximum possible number of distinct outcomes from a given query can be $d + 2$, where $d$ outcomes correspond to the $d$ single transmissions by the $d$ live stations and the $2$ possibilities are when $0$ stations transmit and $\geq 2^{+}$ stations transmit. Thus, for any querying scheme $T$ there must exist a sequence of $\lg_{d+2} \binom{n}{d} \geq \lg_{d+2} (n/d)^d \geq d \lg_{d+2} n - d \in \Omega(d \frac{\lg_2 n}{\lg_2 d})$ queries in the worst case.

\subsection{The new lower bound}
\label{subsec:new}
  We present a new lower bound of $\Omega(d \lg_2 n)$ transmission rounds for deterministic, adaptive conflict resolution strategies in multiple access channels.

  First note that in the lower bound of $\Omega(d \lg_d n)$ rounds sketched above, the gap with respect to the upper bound appears because we grossly estimate that each combinatorial query can result in $d + 2$ possible answers from each query. However, in reality re-appearance of any station, that has already made an individual transmission once may not necessarily give substantially new information to the algorithm. In other words, there may be a way to modify the optimal algorithm so that it does not ask such queries and yet does not have poorer worst case behavior. We establish this fact in Lemma \ref{transformation}.

  Consider an optimal algorithm \textbf{A} for the Q-game and an adversary \textbf{Adv} which makes \textbf{A} run for the longest duration in the worst case (i.e. the worst case adversary for \textbf{A}). Let $T$ be a directed rooted tree that represents the entire sequence of queries made by \textbf{A} and answered by \textbf{Adv} for all possible subsets of live stations of size $d$, as defined in  \cite{GT}: A node $X$ in tree $T$ is associated with some query posed by the algorithm \textbf{A} at that juncture and the outgoing directed edges from $X$ are associated with plausible answers: $0$, $2^{+}$ or the identity of the station that successfully transmitted, referred to as \textbf{id-i}.

	A query in which a station is able to transmit its message successfully, without colliding with others, is called an \textbf{identity-transmitting} query. Throughout, \textbf{S} shall be used to denote the subset of live stations each of which must transmit their messages on MAC singly. Any correct algorithm must ensure at least one \textit{identity-transmitting} query for each station in $S$.

\begin{lemma1}
\label{transformation}
  Consider the Q-game tree $T$ \cite{GT} for algorithm \textbf{A}. Then, there exists an algorithm \textbf{A}' and corresponding Q-game tree $T'$ for which the following conditions hold true:
	\begin{enumerate}
	\item{} The length of any path $p$ from the root to a leaf in $T'$ is less than or equal to the length of the longest path from the root node to leaf node in $T$.
	\item{} There are exactly $d$ \textit{identity-transmitting} edges on every path from the root to a leaf node in $T'$.
	\item{} The number of distinct leaf nodes in tree $T'$ is exactly $\binom{n}{d}$.
	\end{enumerate}
\end{lemma1}

\begin{proof1}
  We execute the following two steps on the Q-game tree $T'=T$ an arbitrary number of times in the process morphing it to Q-game tree $T''$:
(The tree $T'$ referred to in the steps below is the intermediate tree obtained from the previous iteration of applying the two steps).

	\begin{enumerate}
	\item{Removing replications of \textit{identity-transmitting} edges along paths:} Consider any path $p$ in tree $T'$ in which some \textit{identity-transmitting} edge repeats itself. That is, there exist some directed edges $e_i$ and $e_j$, outgoing from nodes $n_i$ and $n_j$ respectively, along path $p$ for which the corresponding transmission sets $Q_i$ and $Q_j$ cause transmissions from the same station $s$. At node $n_j$, the query $Q_j$ is changed to $Q_j - \{s\}$. The answer $0$ now corresponds to the outgoing edge $e_j$ for which the answer in original tree was $\{s\}$. The resulting query $Q_j - \{s\}$ may cause one or more sibling edges of $e_j$ (outgoing from $n_j$) to be \textit{identity-transmitting}. Clearly, these edges cannot be repeating transmissions from stations that have already transmitted along the path from the root to $n_j$. Also, its easy to verify that the only other sibling edge of $e_j$ in tree $T'$ could correspond to answer $2^{+}$. Let it be connected to node $n_k$ in tree $T$ i.e., $e_j = (n_j, n_k)$. The different query $Q_j - \{s\}$ asked at node $n_j$ could result in identity transmissions from certain new stations $s_1, s_2, \dots$ belonging to set $Q_j - \{s\}$; each of these would correspond to a unique edge that is now connected to a replication of the entire sub-tree rooted at node $n_k$ in tree $T$.
	
	\item{Truncating extensions of completed paths:} For every node $n_r$ in the tree, the path from root to node $n_r$ on which all stations have transmitted once, is truncated after the last transmission. This is because the algorithm need not continue after all $d$ stations have singly transmitted their messages.
	Invariably, creating a new \textit{identity-transmitting} edge at a higher position by applying (1) shall always imply that these \textit{identity-transmitting} edges will repeat later in every path in the sub-tree hanging at these edges. This is because in tree $T'$ every path from $n_j$ to a leaf node must have \textbf{identity-transmitting} edges for every station in set $S$.
  \end{enumerate}

  Clearly, the above process must terminate after after a finite number of times after which it cannot be applied any more. If we were to repeat the above process starting from the root node for longer and longer paths, then the process could be repeated at most once for every edge in the tree. Hence, the process needs to be repeated only a finite number of times till none of the paths in the final tree $T''$ has some same station $s$ transmitting twice along it. At the termination of the process, let the Q-game tree obtained be denoted by $T''$. It has the following properties:
		\begin{enumerate}
		\item{}	The length of any path $p$ from the root node to a leaf node in tree $T''$ is less than or equal to the length of the longest path from the root node to the leaf node in tree $T''$. This follows from the observation that the process never elongates any path in the original tree $T''$.

		\item{} It is easy to check that the number of distinct leaf nodes in the tree $T''$ is exactly equal to $\binom{n}{d}$. By the correctness of intermediate Q-game trees in the above process it follows that each of $\binom{n}{d}$ possibilities must exist as a leaf node. Now, suppose that there are $\ge \binom{n}{d}$ leaf nodes in tree $T''$. Let leaf nodes $l_i$ and $l_j$ correspond to the same set of $d$ stations that have their identities transmitted along them. Let $n_r$ be the first common node between the paths $p_i$ from leaf node $l_i$ to root and path $p_j$ from leaf node $l_j$ to root. Then, for the same set $S$ of live stations and query $Q_r$ asked at node $n_r$, there cannot be two different answers. This would violate the validity of algorithm \textbf{A} or the truncation step (in above process) could be applied on tree $T''$, which is a contradiction in either case.
		\end{enumerate}
\end{proof1}

  To all the edges in the tree $T''$, obtained from Claim \ref{transformation}, we assign one of the two colors:
	\begin{itemize}
	\item[] \textbf{Red edge} A directed edge which corresponds to a query to which answer is $0$ (corresponding to silence) or $2^{+}$ (collison detected) is assigned \textbf{Red} color.
	\item[] \textbf{Black edge} A directed edge which corresponds to a query in which a single station transmits is assigned \textbf{Black} color. Such an edge corresponds to an \textit{identity-transmitting} query.
	\end{itemize}

\begin{lemma1}
\label{longest}
  Let $T'$ be a Q-game tree obtained from Claim \ref{transformation}. Let the length of the longest path in tree $T'$ be $x$. Then, the number paths in the tree $T''$ of length $\leq x$ which have exactly $d$ \textit{Black} edges is less than or equal to $\sum_{h=d}^{h=x} \binom{h}{d} 2^{h-d} d!$.
\end{lemma1}

\begin{proof1}
  Let a path have $h$ edges, $d$ of which are \textit{Black}. Then, of the $h$ ordered edges $\binom{h}{d}$ is the number of ways of choosing $d$ edges which are going to be \textit{identity-transmitting} edges. Furthermore, the $d$ \textit{identity-transmitting} edges could themselves be ordered in $d!$ ways. There can be $2^{h-d}$ \textit{Red} edges.

	Thus, the total nummber of paths from the root to the leaf nodes in tree $T''$ can be less than or equal to $\sum_{h=d}^{h=x} \binom{h}{d} 2^{h-d} d!$.
\end{proof1}

  We are now ready to prove a lower bound on the length of the longest path in tree $T''$.

\begin{lemma1}
\label{length}
  There exists a path of length at least $\Omega(d \lg_2 n)$ in tree $T''$.
\end{lemma1}
\begin{proof1}
  From Claim \ref{longest}, the number of paths in tree $T''$ is less than $\sum_{h = d}^{h = x} \binom{h}{d} 2^{h - d} d^d$. This can be simplified to be $\leq \sum_{h = d}^{h = x} \binom{x}{d} 2^{h - d} d^d = \binom{x}{d} 2^{x + 1 - d} d^d$.

  We know that the number of paths from the root to the leaf node in tree $T''$ is exactly equal to $\binom{n}{d}$. Therefore, $\binom{n}{d} \leq \binom{x}{d} 2^{x + 1 - d} d!$. This can be simplified as follows:

	$(\frac{n}{d})^d \leq \binom{n}{d} \leq \binom{x}{d} 2^{x + 1 - d} d^d \leq (\frac{x e}{d})^d 2^{x + 1 - d}$.\\

  Taking logarithm to the base $2$ we have that
    $d \lg_2 n - d \lg_2 d \leq x + 1 - d + d \lg_2 x + d \lg_2 e$\\

  Rearranging the terms we have
  	$x + d \lg_2 x \geq d \lg_2 n - d \lg_2 d - d \lg_2 e - d + 1$\\

  It is easy to verify that this inequality holds true only for $x \in \Omega(d \lg_2 n)$.  
\end{proof1}

  We have the following result.
\begin{theorem1}
\label{finaltheorem}
  Every deterministic, adaptive algorithm for Q-game must require $\Omega(d \lg_2 n)$ transmissions in the worst case.
\end{theorem1}

  The case when the number of live stations may be $\leq d$ can be handled in the same way. Also, it is easy to establish that same asymptotic bound can be achieved when all the stations are not actually required to transmit, but are to be merely identified.

\section*{Acknowledgements} I'd like to thank Darek Kowalski for promptly answering a question about the status of the problem and his interest in it. I thank Kunal Chawla for a few useful discussions on information theoretically greedy adversaries for Q-game.


\end{document}